\documentclass[twocolumn]{svjour3}         
\smartqed  
\usepackage{graphicx}
\usepackage{mathptmx}      
\journalname{Astrophysics and Space Science (ApSS)}
\begin{document}

\title{Hot Cores in the submm - obscured by dust?
}


\author{Rainer Rolffs \and Peter Schilke \and Claudia Comito \and Carolin Hieret \and Friedrich Wyrowski}


\institute{All authors:
 \at Max-Planck-Institut f\"ur Radioastronomie, Bonn          \\
              \email{rrolffs,schilke,ccomito,chieret,wyrowski@mpifr-bonn.mpg.de}           
}

\date{Received: date / Accepted: date}

\maketitle

\begin{abstract}
We present APEX observations of HCN (9-8) and (4-3) lines toward a sample of hot cores. The spectral shapes of the main transitions are asymmetric and self-absorbed, as expected for high optical depth in a possibly infalling envelope. For spherical symmetry, the large column densities of these sources would mean that the central region is obscured by dust above a certain frequency. However, we detected the vibrationally excited satellite lines ($v_2=1$; $J=$9-8) at 797 GHz, which  originate from the inner regions. This indicates that high-frequency ALMA observations of hot core centers will be feasible.


\end{abstract}

\section{Introduction}
\label{intro}
The early stages of star formation are known to occur deep inside interstellar clouds of dust and molecular gas. This is particularly true when massive protostars are produced: the hydrogen-burning reactions in a high-mass protostar are ignited while the object is still accreting and deeply embedded in its parent cloud.  Hence, due to its much higher energy output, it stirs and heats up the natal core in a much more effective way than a low-mass object, giving rise to a hot molecular core.

\begin{figure}
  \includegraphics[width=0.45\textwidth]{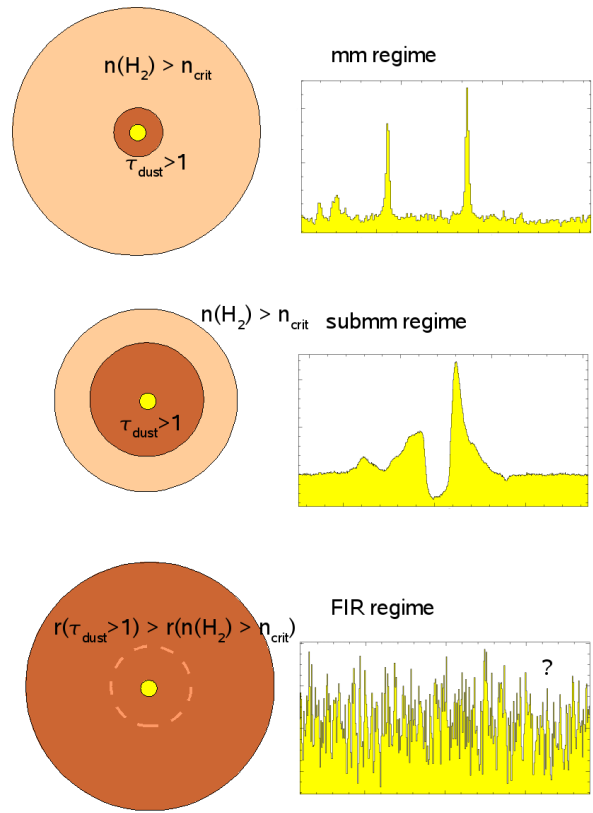}
\caption{In the top panel, the region with high dust opacity (dark brown) is very small, while the region with a density high enough to excite molecules (light brown) is large, one sees emission spectra.  As one goes higher in frequency (and excitation energy), the dust photosphere grows while the critical density region shrinks, and absorption components toward the opaque dust core appear (center panel), while at very high energies and frequencies the region where the molecule can emit becomes embedded in the opaque dust, and no line radiation can escape to be detected by us (bottom panel).}
\label{fig:dust}       
\end{figure}

\begin{figure*}
  \includegraphics[angle=-90, width=0.85\textwidth]{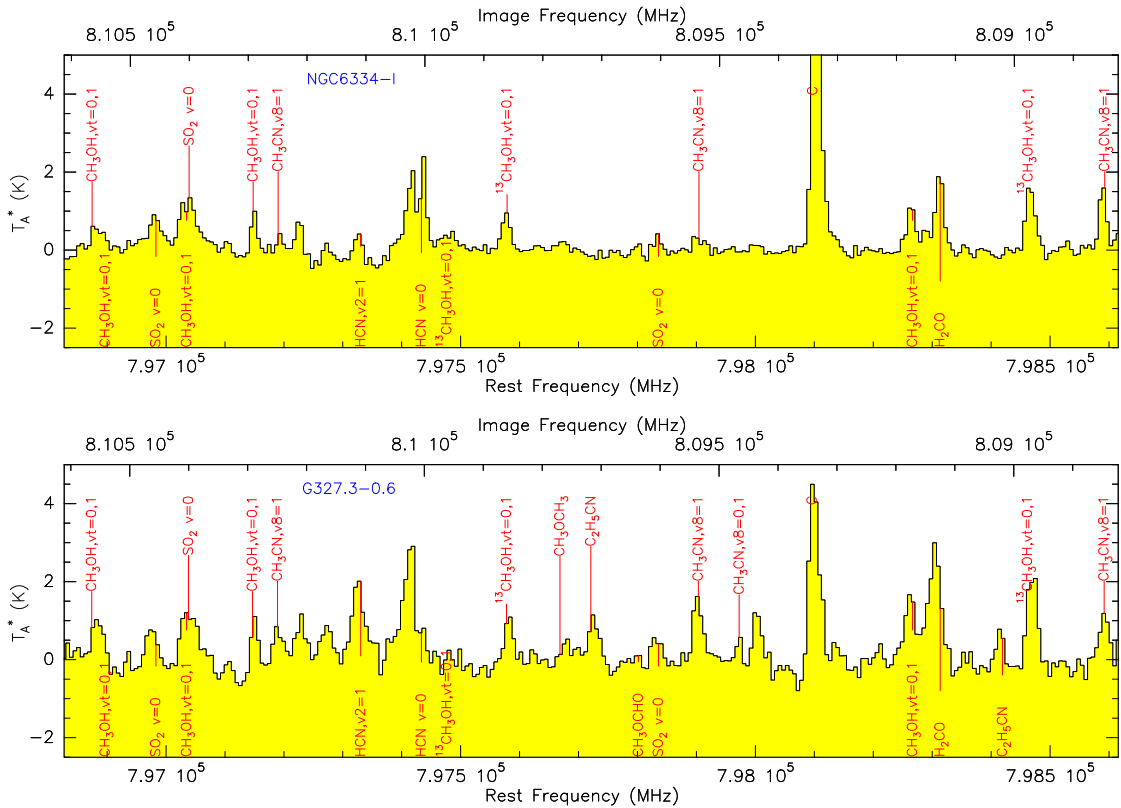}
\caption{The observed spectra around 798/810 GHz are shown for two sources, together with preliminary line identifications}
\label{fig:spec798}       
\end{figure*}

This stage of high-mass star formation is characterized by processes that occur in the dense and heated (to $>$ 100 K) surroundings of the protostar: Ice mantles around dust grains evaporate and the frozen molecules are thereby released into the gas phase. Due to the high temperature and density, the molecules are easily excited, and due to the large column densities, they can readily be observed. Since hot cores display a plethora of molecular lines (see Figure~\ref{fig:spec798}), they can be studied very well using molecular lines, particularly of highly excited transitions (for a review, see e.g. van der Tak 2004 \cite{vdTak04}).

\section{Dust Opacity and Hot Core Geometry}
\label{dust}

Although molecular line studies of hot cores are quite popular at mm and lower submm wavelengths  (e.g. Hatchell et al. 1998 \cite{Hatchell98}), a systematic study of all submm atmospheric windows has, so far, only been carried out for Orion-KL with the CSO  (350 and 650 GHz, Schilke et al. 1997 \cite{Schilke97}, 2001 \cite{Schilke01}; 850 GHz, Comito et al. 2005 \cite{Comito05}).

Often considered a template for the study of the hot-core phase of massive
star formation, Orion-KL is in fact not representative at all:
\begin{itemize} 
\item  The mass involved, 10 $M_\odot$,  is about 2 orders of magnitude lower than the mass of many other massive hot cores (e.g. G10.47, Olmi et al. 1996 \cite{Olmi96}), and its extremely strong molecular lines are due mostly to its proximity (i.e., to the relatively large beam filling factors); 
\item Its anomalous geometry (cf.  the ``blister'' model, Genzel \& Stutzki 1989 \cite{Genzel89}) may result in an abnormally low dust opacity along the line of sight, a condition that is not representative of ``typical'' hot-core sources. 
\end{itemize}

In fact, for a typical H$_2$ column density of $10^{25}$ cm$^{-2}$, {\bf the dust opacity
$\tau_{dust} \propto \nu^{\beta}$, with $\beta \approx 1.5$, will become $>$ 1 at a frequency of about 550 GHz}. Thus, while Orion does look similar to many other hot cores at frequencies of $\approx 350$ GHz, an extrapolation to shorter wavelengths is by no means obvious (see Figure~\ref{fig:dust}).

Molecular observations of massive ($>$ 100 M$_\odot$) cores in the 690 or 850 GHz (450 or 350 $\mu$m) band are thus likely to show not only quantitative but also qualitative differences with respect to lower-frequency observations. At least two scenarios may be encountered:

 (i) a geometrically straightforward one, where the observed object is spherical in shape and isotropic in density structure. In this case, the high dust opacity at 350/450 $\mu$m  will obscure the innermost, hottest and densest regions of the hot molecular core, while opening the path to absorption-line studies of the outer envelope.

 (ii) one in which the hot-core component is actually made up by a cluster of high mass clumps. Here, in spite of the high nominal  H$_2$ column densities (hence high $\tau_{dust}$), the radiation from close to the central object(s) could still be observable, since the radiation can leak out between the clumps. The same applies to many non-spherical geometries  (e.g. face-on disks, hollow outflow cones).

Scenario (i) above would be bad news for high-frequency ALMA observations, but it would be fatal for planned high frequency Herschel/HIFI surveys, which will extend to even higher frequencies, 1.9 THz, corresponding to a wavelength of 150 $\mu$m.  Of course, one would not lose all spectral features, since low excitation lines would still show up in absorption and would provide valuable information (similar to the IR absorption lines), but the direct view to the central engine would be blocked.
\begin{figure*}
  \includegraphics[angle=-90, width=0.85\textwidth]{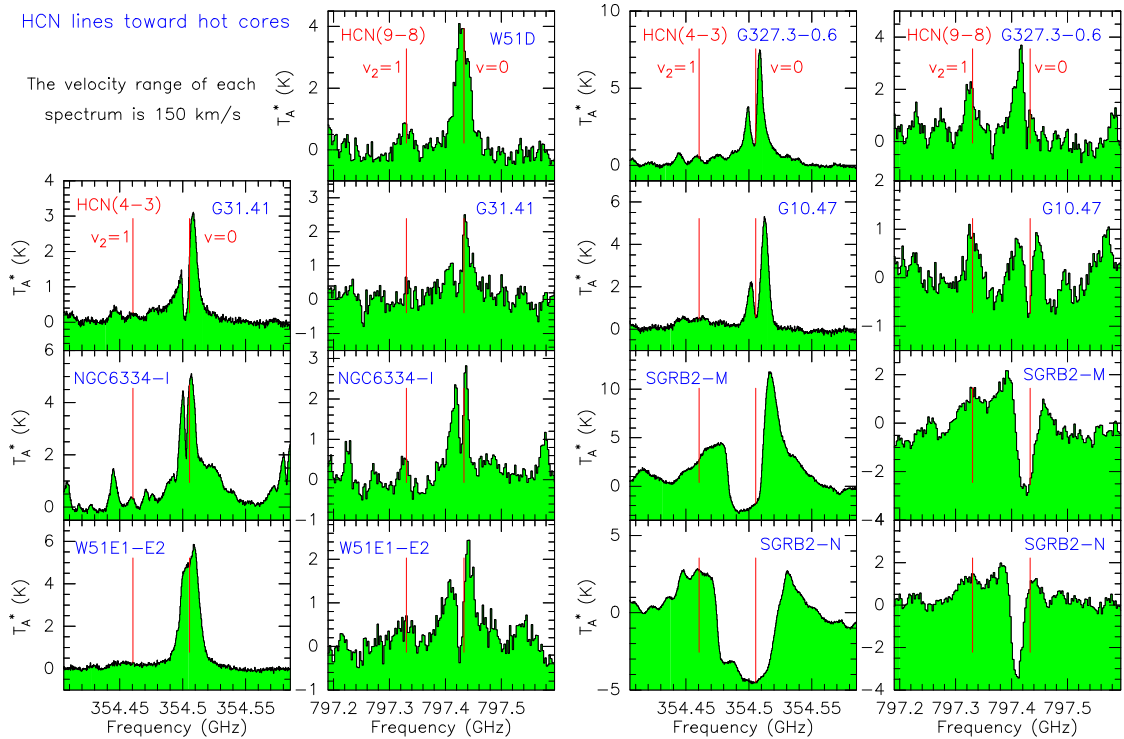}
\caption{The observed HCN 9-8 and 4-3, $v_2$=0 and 1 lines are shown. The displayed spectral parts cover 150 km/s in velocity. }
\label{fig:hcnlines}       
\end{figure*}

\section{Observations}
\label{obs}

To discern between the proposed scenarios, and to help selecting suitable objects for the Herschel/HIFI spectral scan key project, we observed the vibrationally excited HCN(9-8, $v_2=1$) line at 797 GHz. Because of its high critical density ($>10^{10}$ cm$^{-3}$), this transition can hardly be excited by collisions, but can be pumped by 14 $\mu$m IR radiation (corresponding to the the energy of the $v_2=1$  state) which is emitted by warm ($>$ 150 K) dust (Carroll \& Goldsmith 1981 \cite{Carroll81}, Hauschildt et al. 1993 \cite{Hauschildt93}). It therefore comes from either unusually dense and warm clumps or from regions of a high IR radiation field, i.e. the line originates most likely from close to the central star.

Using the FLASH (Heyminck et al. 2006 \cite{Heyminck06}) and APEX-2A (Risacher et al. 2006 \cite{Risacher06}) instruments on APEX (G\"usten et al. 2006 \cite{Gusten06}), we observed the $J$=9-8 and 4-3 transitions in the $v_2$=0 and 1 vibrational states, as well as the 4-3 transition of H$^{13}$CN and HC$^{15}$N. Using two fast fourier transform spectrometers (Klein et al. 2006 \cite{Klein06}) as backends, three 1.8 GHz wide bands were covered, centered around 798/810, 342/354 and 345/357 GHz (see Figure~\ref{fig:spec798} as an example). The source sample consists of 10 hot cores (see Figure~\ref{fig:hcnlines} plus 17233-3606 and W33A). 

The HCN(4-3, $v_2=1$) should not suffer significantly from high dust opacities at 850 $\mu$m and can be used to calibrate the intrinsic strength of the lines.

\section{Preliminary Results and Conclusions}
\label{results}
We found evidence for the (9-8, $v_2=1$) line toward all sour- ces where the (4-3, $v_2=1$) was detected, indicating that, at least at this wavelength and for this source sample, the central core is still visible.  Hence ALMA observations of the central engines of hot cores even at the highest frequencies appear to be feasible, although the cores may shut down at higher frequencies relevant for Herschel/HIFI.  

The ground state lines often display self-absorption features. Their asymmetric line shapes with the blue part being the stronger one could be indicative of infall motion. To extract the information on the source structure from the line strengths and shapes, we plan to analyze the data with detailed radiative transfer modeling (Hogerheijde \& van der Tak 2000 \cite{Hogerheijde00}). Preliminary results indicate that modelling the source as a spherical cloud cannot reproduce all observed lines, indicating deviations from this simple structure. Investigations of effects of such deviations are under way.


%




\end{document}